\documentclass[twocolumn,showpacs,aps,floatfix,pra,superscriptaddress,nofootinbib]{revtex4}
\usepackage{color}
\usepackage{graphicx}
\usepackage{bm}
\usepackage{amsmath}
\begin{document}
\title{Quantum Dynamics in a Time-dependent Hard-Wall Spherical Trap}
\author{S. V. Mousavi}
\email{vmousavi@qom.ac.ir}
\affiliation{Department of Physics, The University of Qom, P. O. Box 37165, Qom, Iran}

\begin{abstract}

Exact solution of the Schr\"{o}dinger equation is given for a
particle inside a hard sphere whose wall is moving
with a constant velocity. Numerical computations are presented for
both contracting and expanding spheres. The
propagator is constructed and compared with the propagator of a
particle in an infinite square well with one wall in
uniform motion.

\end{abstract}
\maketitle
%

\section{Introduction}

Solving the Schr\"{o}dinger equation with time-dependent boundary
conditions, including moving ones, is a very hard work and can only
be done in a few cases. See, {\it e.g.,}
\cite{Ma-JPA-1992, MaDe-PLA-1991, DoRi-AJP-1969} and
\cite{CaCaMu-PR-2009} for a recent review.
Very interesting effects are seen in such problems; diffraction in
time \cite{Mo-PR-1952} is just such an instance. This phenomenon is
characterized by quantum temporal oscillations in matter waves
released from a confining region.
It was shown when a wall acting as a perfect mirror,
moves with finite velocity along the direction of propagation of a
beam, the visibility of the fringes is enhanced
\cite{CaMuCl-PRA-2008}.
Moshinsky's theoretical work has been extended to the case of
particles which are suddenly released from a 1D box
\cite{Go-PRA-2002_CaMu-EP-2006_Mo-JPA-2010} and to particles with
angular momentum \cite{Go-PRA-2003} initially trapped in a hard
spherical box.
Exact solutions of the Schr\"{o}dinger equation for a particle in a
1D box with a moving wall have been found \cite{MaDe-PLA-1991,
DoRi-AJP-1969}. Using the semiclassical approximation, Luz and Cheng
\cite{LuCh-JPA-1992} evaluated the exact propagator of the problem.
Grosche \cite{Gr-PLA-1993} did this task independently by means of
an exact (mid-point) summation of the perturbation series with
point-like perturbations.

The motivation of the present work is that sudden removal of the
boundary is an idealized case, thus we consider a moving boundary
instead of a sudden removal of the boundary. The limit of infinite
velocity of the moving boundary clearly corresponds to the sudden
removal case. Possible applications to optical effects
connected with moving mirrors are additionally encouraging. For
these reasons we aim to solve the Schr\"{o}dinger equation for a
particle in a hard sphere with varying radius.

\section{Exact solution}

Consider a particle with mass $\mu$ inside a hard
sphere with a time-dependent radius $L(t)$. The potential energy
function is zero if $r < L(t)$ and infinite otherwise. The
Schr\"{o}dinger equation is then
\begin{eqnarray} \label{eq: sch}
i \hbar \frac{\partial}{\partial t} \Psi({\bf{r}}, t) &=& -\frac{\hbar^2}{2 \mu} \nabla^2 \Psi({\bf{r}}, t)~,
\end{eqnarray}
with the boundary condition $\Psi({\bf{r}}, t)|_{r=L(t)} = 0$.

The {\it instantaneous} energy eigenfunctions and eigenvalues are respectively
\begin{eqnarray} \label{eq: instan-waves}
u_{l n m}({\bf{r}}, t) &=& \sqrt{\frac{2}{L^3(t)}} ~ \frac{1}{| j_{l + 1}(x_{l n}) |} ~ j_{l} \left( x_{l n}  \frac{r}{L(t)} \right) Y_{l m}(\theta, \phi)~,\\
E_{l n}(t) &=& \frac{\hbar^2 x^2_{l n}}{2\mu L^2(t)}~,
\end{eqnarray}
$l = 0, 1, 2, ...$; $n = 1, 2, 3, ...$ and $m = -l, -l + 1, ...,
l-1, l$, where $j_{l}(x)$ and $Y_{l m}(\theta, \phi)$ are
respectively spherical Bessel functions and harmonics. $x_{l n}$ is
the $n^{\text{th}}$ zero of the spherical Bessel function of order
$l$, {\it i.e.,} $j_{l}(x_{l n}) = 0$. It must be noted that all
Bessel functions with $l \neq 0$ have a zero at the origin,
but to have a non-zero wave function these zeros
must be excluded.

Using the method of "separation of variables" for solving the partial differential equation (\ref{eq: sch}), we propose the solution
\begin{eqnarray} \label{eq: prosol}
\Psi({\bf{r}}, t) &=& \frac{U(r, t)}{r}Y_{l m}(\theta, \phi)~,
\end{eqnarray}
where we have used the spherical symmetry of the Hamiltonian.

Putting eq. (\ref{eq: prosol}) into eq. (\ref{eq: sch}) one gets
\begin{eqnarray} \label{eq: radial_Sch_1}
i \hbar \frac{1}{r} \frac{\partial U(r, t)}{\partial t} = -\frac{\hbar^2}{2 \mu}
\left[ \frac{1}{r} \frac{\partial^2 U(r, t)}{\partial r^2} - \frac{l(l +1)}{r^2}  \frac{U(r, t)}{r} \right]~.
\end{eqnarray}
The radial part of the proposed wave-function, $R(r, t) = U(r, t)/r$, must be zero on the shell, thus the boundary conditions on $U(r, t)$ are
$U(r, t)|_{r=0} = 0 = U(r, t)|_{r=L(t)}$.

Now, we follow \cite{MaDe-PLA-1991} to solve the eq. (\ref{eq: radial_Sch_1}). By defining a new coordinate
\begin{eqnarray} \label{eq: indep_change_1}
s = \frac{r}{L(t)} ~,
\end{eqnarray}
we get
\begin{eqnarray} \label{eq: radial_Sch_2}
i \hbar \frac{\partial U(s, t)}{\partial t} &=& i\hbar \frac{\dot{L}(t)}{L(t)} s \frac{\partial U(s, t)}{\partial s}
- \frac{\hbar^2}{2 \mu}  \frac{1}{L^2(t)} \nonumber \\
&& \times
\left[ \frac{\partial^2 U(s, t)}{\partial s^2} - \frac{l(l +1)}{s^2} {U(s, t)} \right]~,
\end{eqnarray}
where $\dot{L}(t) = dL(t)/dt$ and moving boundary conditions are replaced by fixed-boundary ones; $U(s, t)|_{s=0} = 0 = U(s, t)|_{s=1}$.
When the transformation
\begin{eqnarray} \label{eq: dep_change_1}
U(s, t) &=& \sqrt{\frac{2}{L(t)}} \exp \left[ \frac{i\mu}{2\hbar}L(t) \dot{L}(t) s^2 \right]  \varphi(s, t) ~.
\end{eqnarray}
is introduced in eq. (\ref{eq: radial_Sch_2}), one obtains
\begin{eqnarray} \label{eq: radial_Sch_3}
i \hbar \frac{\partial \varphi(s, t)}{\partial t} &=&  -\frac{\hbar^2}{2 \mu} \frac{1}{L^2(t)}
\left[ \frac{\partial^2 \varphi(s, t)}{\partial s^2} - \frac{l(l +1)}{s^2} {\varphi(s, t)} \right]~,
\end{eqnarray}
for the uniform motion of the wall, {\it i.e.,}
$\ddot{L}(t) = 0$. Boundary conditions on $\varphi(s, t)$ are
$\varphi(s, t)|_{s=0} = 0 = \varphi(s, t)|_{s=1}$. Defining the new
time variable $\tau$ as
\begin{eqnarray} \label{eq: indep_change_2}
\tau(t) &=& \int_0^t \frac{d t^{\prime}}{L^2(t^{\prime})}  ~, ~~~
\Longrightarrow ~~ \frac{\partial}{\partial t} = \frac{1}{L^2(t)} \frac{\partial}{\partial \tau}~,
\end{eqnarray}
Eq. (\ref{eq: radial_Sch_3}) transforms to
\begin{eqnarray} \label{eq: radial_Sch_4}
i \hbar \frac{\partial \varphi(s, \tau)}{\partial \tau} &=&  -\frac{\hbar^2}{2 \mu}
\left[ \frac{\partial^2 \varphi(s, \tau)}{\partial s^2} - \frac{l(l +1)}{s^2} \varphi(s, \tau) \right]~.
\end{eqnarray}
Inserting $\varphi(s, \tau) = \exp(-iE^{\prime}\tau/\hbar ) \psi(s)$
in (\ref{eq: radial_Sch_4}), one gets
\begin{eqnarray} \label{eq: radial_Sch_5}
E^{\prime} \psi(s) &=&  -\frac{\hbar^2}{2 \mu} \left[ \frac{\partial^2 \psi(s)}{\partial s^2} - \frac{l(l +1)}{s^2} \psi(s) \right]~.
\end{eqnarray}
By introducing new variable $k^2 = 2\mu E^{\prime}/\hbar^2$, we obtain
\begin{eqnarray} \label{eq: radial_Sch_6}
\frac{\partial^2 \psi(s)}{\partial s^2} + \left( k^2- \frac{l(l +1)}{s^2} \right) \psi(s) &=& 0 ~.
\end{eqnarray}
The solutions of this equation are spherical Bessel functions
\begin{eqnarray} \label{eq: radial_Sch_5}
\psi(s) &=& s[c_1 j_{l}(ks) + c_2 n_{l}(ks)] ~.
\end{eqnarray}
If the radial wave-function $R(r)$ is finite at the origin, $c_2 = 0$. The requirement that $\psi(s) = 0$ at $s=1$ means that $k$ can take on only those
special values
\begin{eqnarray} \label{eq: k_values}
k_{l n} &=& x_{l n} ~~~~~~~~~~~~~~ (n = 1, 2, 3, ...)~.
\end{eqnarray}
%
For the uniform change of the radius with velocity
$u$
\begin{eqnarray} \label{eq: shell}
L(t) &=& a + u t~,
\end{eqnarray}
where $a$ is the initial radius, one has
\begin{eqnarray} \label{eq: tau(t)}
\tau(t) &=& \frac{t}{a (a + ut)}~.
\end{eqnarray}

By using equations (\ref{eq: tau(t)}), (\ref{eq: shell}), (\ref{eq: k_values}), (\ref{eq: dep_change_1}) and (\ref{eq: indep_change_1}) one obtains
\begin{eqnarray} \label{eq: radialwave}
R_{l n} (r, t)&=& c_1 \sqrt{\frac{2}{L(t)}} \exp \left[ \frac{i \mu}{2\hbar} u \frac{r^2}{L(t)} -i\frac{\hbar}{2 \mu} x^2_{l n} \frac{t}{aL(t)} \right]   \nonumber \\
&\times &
j_{l} \left( x_{l n}  \frac{r}{L(t)} \right)   ~,
\end{eqnarray}
for the radial part of the wave-function. Unknown coefficient $c_1$ is determined by the normalization condition
\begin{eqnarray} \label{eq: normalization_1}
\int_0^{L(t)} dr r^2 \int d\Omega |\Psi_{l n m}({\bf{r}}, t)|^2 &=& 1~,
\end{eqnarray}
where
\begin{eqnarray} \label{eq: eigenFunctions}
\Psi_{l n m}({\bf{r}}, t) &=& R_{l n}(r, t) Y_{l m}(\theta, \phi)
\end{eqnarray}
are the solutions of the Schr\"{o}dinger equation (\ref{eq: sch})
for a particle in a spherical box with a wall in
uniform motion and $\int d\Omega = \int_{-1}^{1} d(\cos \theta)
\int_{0}^{2\pi} d\phi$.

Using the orthogonality of the spherical Bessel functions \cite{Arfken-book-2005}
\begin{eqnarray} \label{eq: bessel-ortho}
\int_0^1 ds~s^2~ j_{l}(x_{l n}s) j_{l}(x_{l m}s)   &=& \frac{1}{2} [ j_{l + 1}(x_{l n}) ]^2 \delta_{nm}~,
\end{eqnarray}
one obtains
\begin{eqnarray} \label{eq: norm_coefficient}
|c_1|^2 &=& \frac{1}{L^2(t)} \frac{1}{[ j_{l + 1}(x_{l n}) ]^2}~.
\end{eqnarray}
Thus apart from a phase factor, one obtains
\begin{eqnarray} \label{eq: norm_waves}
\Psi_{l n m}({\bf{r}}, t) &=& \frac{1}{L(t)} \sqrt{\frac{2}{L(t)}} \frac{1}{| j_{l + 1}(x_{l n}) |}
\nonumber \\
& & \times
\exp \left[ \frac{i \mu}{2\hbar} u \frac{r^2}{L(t)} - i\frac{\hbar}{2 \mu} x^2_{l n} \frac{t}{aL(t)} \right]
\nonumber \\
& & \times
j_{l} \left( x_{l n}  \frac{r}{L(t)} \right)
 Y_{l m}(\theta, \phi) \nonumber\\
&\equiv&  \exp \left[ i \alpha \xi(t) \left(\frac{r}{L(t)} \right)^2 - i x^2_{l n} \frac{1-1/\xi(t)}{4 \alpha} \right]
\nonumber \\
& & \times
u_{l n m}({\bf{r}}, t)~,
\end{eqnarray}
where we have introduced new dimensionless parameters $\alpha = \mu a u /(2 \hbar)$ and $\xi(t) = L(t)/a$.

Functions $\Psi_{l n m}({\bf{r}}, t)$ vanish at $r = L(t)$, remain
normalized as the radius changes, and form a complete
orthogonal set. The general solution of eq. (\ref{eq: sch}) is a
superposition of functions (\ref{eq: norm_waves})
\begin{eqnarray} \label{eq: geral-sol_1}
\Psi({\bf{r}}, t) &=&
\sum_{l^{\prime}=0}^{\infty} \sum_{n^{\prime}=1}^{\infty} \sum_{m^{\prime} = -l^{\prime}}^{l^{\prime}}
c_{l^{\prime} n^{\prime} m^{\prime}}\Psi_{l^{\prime} n^{\prime} m^{\prime}}({\bf{r}}, t)~,
\end{eqnarray}
with time-independent coefficients $c_{l^{\prime} n^{\prime} m^{\prime}}$ determined from the relation
\begin{eqnarray} \label{eq: coef_1}
c_{l^{\prime} n^{\prime} m^{\prime}} = \int_0^a dr ~ r^2 \int d\Omega ~
\Psi^*_{l^{\prime} n^{\prime} m^{\prime}}({\bf{r}}, 0) \Psi({\bf{r}}, 0) ~,
\end{eqnarray}

General solution can also be expanded in terms of instantaneous eigenfunctions as
\begin{eqnarray} \label{eq: geral-sol_2}
\Psi({\bf{r}}, t) &=&
\sum_{l^{\prime}=0}^{\infty} \sum_{n^{\prime}=1}^{\infty} \sum_{m^{\prime} = -l^{\prime}}^{l^{\prime}}
b_{l^{\prime} n^{\prime} m^{\prime}}(t)
u_{l^{\prime} n^{\prime} m^{\prime}}({\bf{r}}, t)~,
\end{eqnarray}
now with time-dependent coefficients $b_{l^{\prime} n^{\prime} m^{\prime}}(t)$ determined from the relation
\begin{eqnarray} \label{eq: coef_2}
b_{l^{\prime} n^{\prime} m^{\prime}}(t) = \int_0^{L(t)} dr~r^2 \int d\Omega ~
u^*_{l^{\prime} n^{\prime} m^{\prime}}({\bf{r}}, t) \Psi({\bf{r}}, t) ~,
\end{eqnarray}
Using eqs. (\ref{eq: coef_2}) and (\ref{eq: geral-sol_1}) and the orthogonality of spherical harmonics, one finds
\begin{eqnarray} \label{eq: relation_coefs}
b_{l^{\prime} n^{\prime} m^{\prime}}(t) &=& \frac{2}{| j_{l^{\prime} + 1}(x_{l^{\prime} n^{\prime}}) |}
\sum_{n^{\prime \prime}=1}^{\infty} c_{l^{\prime} n^{\prime \prime} m^{\prime}}
\frac{1}{| j_{l^{\prime} + 1}(x_{l^{\prime} n^{\prime \prime}}) |}
\nonumber \\
& & \times
\exp \left[ -i x^2_{l^{\prime} n^{\prime\prime}} \frac{1-1/\xi(t)}{4 \alpha} \right]
I^*_{l^{\prime} n^{\prime} n^{\prime \prime}}(t, \alpha)~,
\end{eqnarray}
where
\begin{eqnarray} \label{eq: integral}
I_{l^{\prime} n^{\prime} n^{\prime \prime}}(t, \alpha) =
\int_0^1 ds ~ s^2 e^{-i\alpha \xi(t) s^2} j_{l^{\prime}}(x_{l^{\prime} n^{\prime}} s) j_{l^{\prime}}(x_{l^{\prime} n^{\prime \prime}} s)~.
\end{eqnarray}
This integral is not elementary and following the procedure of
\cite{Go-PRA-2003}, can be reduced to a combination of terms
involving the Ferensel integrals and derivative of Legendre
polynomials.

The expectation value of the energy of the particle is obtained from
\begin{eqnarray} \label{eq: Energy_exp}
\langle E(t) \rangle  &=&
\sum_{l^{\prime} n^{\prime} m^{\prime}}  |b_{l^{\prime} n^{\prime} m^{\prime}}(t)|^2 ~ E_{l^{\prime} n^{\prime} m^{\prime}}(t) ~.
\end{eqnarray}

If the particle is initially in an energy eigenstate, {\it i.e.,} $\Psi({\bf{r}}, 0) = u_{l n m}({\bf{r}}, 0)$,
then
\begin{eqnarray} \label{eq: c_cof_eigen}
c_{l^{\prime} n^{\prime} m^{\prime}} = \delta_{l l^{\prime}}\delta_{m m^{\prime}}
\frac{2}{| j_{l + 1}(x_{l n}) | | j_{l + 1}(x_{l n^{\prime}}) |} I_{l n n^{\prime}}(0, \alpha)~,
\end{eqnarray}
which is not an unexpected result as quantum numbers $l$ and $m$ do not change.

\section{numerical calculations}

\begin{figure}
\centering
\includegraphics[width=7cm,angle=-90]{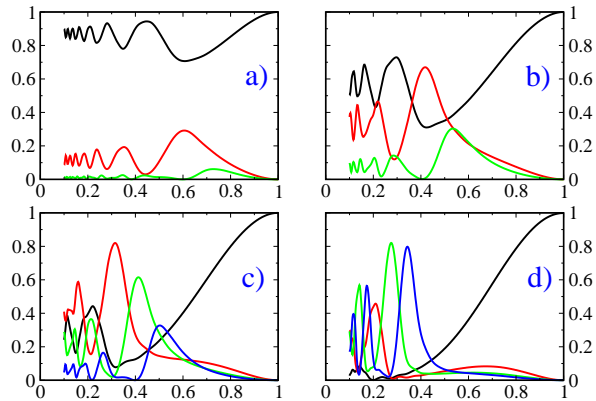}
\caption{(Color online) Transition probabilities versus $\xi(t)$ for different values of velocity parameter $\alpha$: a) $\alpha = -2$;
b) $\alpha = -4$; c) $\alpha = -6$ and d) $\alpha = -10$. In each part the black curve shows $|b_{11m}|^2$, red one $|b_{12m}|^2$, green one $|b_{13m}|^2$ and the blue one $b_{14m}|^2$.}
\vspace*{0.5cm}
\label{fig: B_Coefficients}
\end{figure}

\begin{figure}
\centering
\includegraphics[width=7cm,angle=-90]{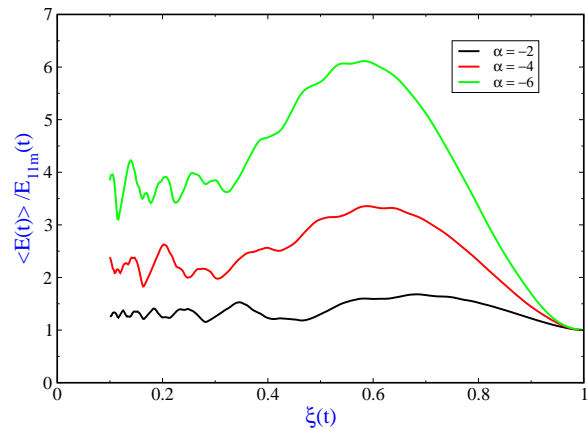}
\caption{(Color online) Ratio of the energy expectation value to the instantaneous first excited energy as a function of $\xi(t)$ for three different values of velocity parameter.}
\vspace*{0.0cm}
\label{fig: Energy_Expectation}
\end{figure}

Numerical computations are shown in figs. \ref{fig: B_Coefficients}
and \ref{fig: Energy_Expectation} for a particle that is initially
in the first excited state with three fold degeneracy. In this case
we have
\begin{eqnarray} \label{eq: Energy_exp_2}
\frac{\langle E(t) \rangle}{E_{11m}(t)}  &=&
\sum_{n^{\prime}}  |b_{1 n^{\prime} m}(t)|^2 ~ \left( \frac{x_{1 n^{\prime}}}{x_{11}} \right)^2~.
\end{eqnarray}
for the ratio of energy expectation value to the instantaneous first
excited state energy.

Figure \ref{fig: B_Coefficients} shows the squares of energy
eigenfunction expansion coefficients versus $\xi(t)$ for three
different contraction rates $\alpha$. For these
values of $\alpha$, it was found that series (\ref{eq:
relation_coefs}) converges for the first ten terms.

Figure \ref{fig: Energy_Expectation} shows the ratio of the
expectation value of the energy to the energy the particle would
have if it remained in the first excited state $u_{11m}$
for the sphere in contraction. Here
fifteen terms in eq. (\ref{eq: Energy_exp_2}) leads to convergency.

\begin{figure}
\centering
\includegraphics[width=7cm,angle=-90]{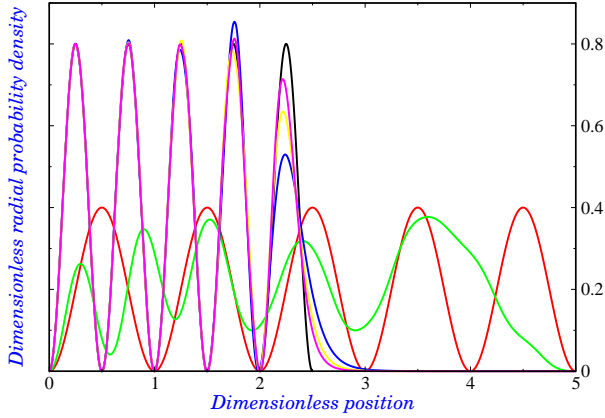}
\caption{(Color online) Dimensionless radial probability density
$\rho_{l n}(\eta_{l n}, T_{l n})$ for a particle initially in the
state $u_{0,5,0}$, against dimensionless position coordinate
$\eta_{l n}$ at dimensionless time coordinate $T_{l n}^{(0)}$, for
six different values of expansion rate; $\alpha = 0$ (black curve),
$\alpha = 0.01 \alpha_{l n}$ (red curve), $\alpha = \alpha_{l n}$
(green curve), $\alpha = 10 \alpha_{l n}$ (blue curve), $\alpha = 15
\alpha_{l n}$ (yellow curve) and  $\alpha = 20 \alpha_{l n}$
(magenta curve); where $\alpha_{l n} = x_{l n}/2$.} \vspace*{0.5cm}
\label{fig: wave_r}
\end{figure}

We have plotted dimensionless radial probability density $\rho_{l
n}(\eta_{l n}, T_{l n}) = \lambda_{l n}^3 \eta_{l n}^2 |R(\eta_{l
n}, T_{l n})|^2$ in fig. \ref{fig: wave_r} for a particle initially
in the state $u_{0,5,0}$, against dimensionless position coordinate
$\eta_{l n} = r/\lambda_{l n}$ at dimensionless time coordinate
$T_{l n}^{(0)} = \nu_{l n} t^{(0)}$ and in fig. \ref{fig: wave_t}
for a particle initially in the state (a) $u_{0,15,0}$ and (b)
$u_{0,100,0}$, against dimensionless time coordinate $T_{l n} =
\nu_{l n} t$ at dimensionless observation point  $\eta_{l n}^{(0)} =
r^{(0)}/\lambda_{l n}$;  where $\lambda_{l n} = 2\pi a/x_{l n}$ and
$\nu_{l n} = E_{l n}/h$. In our calculations $r^{(0)} = 2a$ and
$t^{(0)} = (r^{(0)}-a)/u$.

$T_1$ and $T_2$ are dimensionless classical flight times from the
front and back edges of the sphere to the dimensionless observation
point $\eta_{l n}^{(0)}$ for a particle in the state $u_{lnm}$.

\begin{figure}
\centering
\includegraphics[width=7cm,angle=-90]{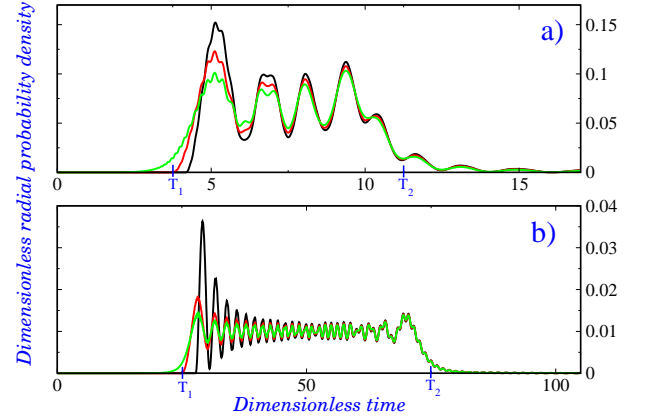}
\caption{(Color online) Dimensionless radial probability density
$\rho_{l n}(\eta_{l n}, T_{l n})$ for a particle initially in the
state (a) $u_{0,15,0}$ and (b) $u_{0,100,0}$, against dimensionless
time coordinate $T_{l n} = \nu_{l n} t$ at dimensionless observation
point $\eta_{l n}^{(0)}$, for three different values of velocity
parameter; $\alpha = 0.9 \alpha_{l n}$ (black curve), $\alpha =
\alpha_{l n}$ (red curve), $\alpha = 2 \alpha_{l n}$ (green curve);
where $\alpha_{l n} = x_{l n}/2$. $T_1$ and $T_2$ are dimensionless
classical flight times from the front and back edges of the sphere
to the dimensionless observation point.} \vspace*{0.5cm} \label{fig:
wave_t}
\end{figure}

\section{Propagator}

One can construct the propagator as follows
\begin{eqnarray*}
|\Psi(t) \rangle &=& S(t, t_0) |\Psi(t_0) \rangle
\nonumber \\
&=&
\sum_{l n m} \sum_{l^{\prime} n^{\prime} m^{\prime}} | \Psi_{l n m}(t) \rangle
\langle \Psi_{l n m}(t) | S(t, t_0) | \Psi_{l^{\prime} n^{\prime} m^{\prime}} (t_0) \rangle
\nonumber \\
& & \times
\langle \Psi_{l^{\prime} n^{\prime} m^{\prime}}(t_0)| \Psi(t_0) \rangle
\nonumber \\
&=&
\sum_{l n m} | \Psi_{l n m}(t) \rangle \langle \Psi_{l n m}(t)| \Psi(t_0) \rangle ~,
\end{eqnarray*}
where $S(t, t_0)$ is the time evolution operator and we have used
the fact that if the particle is in the sate $|
\Psi_{l n m} \rangle $ at $t_0$, it remains in
that state as the wall moves,
{\it{i.e.,}} $S(t, t_0) | \Psi_{l n m}(t_0)\rangle = | \Psi_{l n
m}(t) \rangle $. Now, we write this equation in the form
\begin{eqnarray} \label{eq: wave-function}
\Psi({\bf{r}}, t) = \int_0^a dr^{\prime} {r^{\prime}}^2 \int d\Omega^{\prime} K({\bf{r}}, t ; {\bf{r}}^{\prime}, t^{\prime})
\Psi(r^{\prime}, \theta^{\prime}, \phi^{\prime}, t^{\prime}) ~,
\end{eqnarray}
where we have introduced the propagator as,
\begin{eqnarray} \label{eq: propagator}
K({{\bf{r}}}, t ; {{\bf{r}}}^{\prime}, t^{\prime}) &=& \sum_{l=0}^{\infty} \sum_{n=1}^{\infty} \sum_{m = -l}^{l}
\Psi_{l n m}({\bf{r}}, t) \Psi^*_{l n m}({{\bf{r}}}^{\prime}, t^{\prime}) \nonumber \\
&=&
\frac{2}{L^{3/2}(t)L^{3/2}(t^{\prime})} \sum_{l n m} \frac{1}{[ j_{l + 1}(x_{l n}) ]^2}~
\nonumber \\
& & \times
\exp \left[ \frac{i\mu u}{2\hbar} \left( \frac{r^2}{L(t)} -  \frac{{r^{\prime}}^2}{L(t^{\prime})}  \right) \right]
\nonumber \\
& & \times
\exp \left[ -\frac{i \hbar}{2\mu} \frac{x^2_{l n}}{a}\left( \frac{t}{L(t)} -  \frac{t^{\prime}}{L(t^{\prime})} \right) \right]
\nonumber \\
& & \times
j_{l} \left( x_{l n}  \frac{r}{L(t)} \right)~j_{l} \left( x_{l n}  \frac{r^{\prime}}{L(t^{\prime})} \right)~
\nonumber \\
& & \times
Y_{l m}(\theta, \phi)~Y^*_{l m}(\theta^{\prime}, \phi^{\prime})~.
\end{eqnarray}

One sees when $l = 0$, eq. (\ref{eq: radial_Sch_2}) reduces to eq.
(4) of \cite{MaDe-PLA-1991}, {\it i.e.,} $l = 0$ corresponds to a
particle in a 1D box with the left wall at $x=0$ and the right wall
in uniform motion. In order to have the relation
\begin{eqnarray}
\Psi(x, t) = \int_0^a K_{\text{1D}}(x, t; x^{\prime}, 0) \Psi(x^{\prime}, 0) dx^{\prime}~,
\end{eqnarray}
in 1D, we must write 1D propagator as
\begin{eqnarray} \label{eq: 1d-3d-prop}
K_{\text{1D}}(x, t ; x^{\prime}, t^{\prime}) &=& \frac{r r^{\prime}}{4\pi} K(r, t ; r^{\prime}, t^{\prime})
\nonumber \\
&\equiv&
\sum_{n=1}^{\infty}
\frac{U_{0n}(r)}{\sqrt{4\pi}} \frac{U_{0n}(r^{\prime})}{\sqrt{4\pi}}~.
\end{eqnarray}

Preserving just the terms with $l = 0$,
(\ref{eq: propagator}) leads
\begin{eqnarray} \label{eq: 1_1D_prop}
K(r, t ; r^{\prime}, t^{\prime}) &=& \frac{2}{L^{3/2}(t)L^{3/2}(t^{\prime})} \sum_{n=1}^{\infty} \frac{1}{[ j_1(x_{0 n}) ]^2}~
\nonumber \\
& & \times
\exp \left[ \frac{i\mu u}{2\hbar} \left( \frac{r^2}{L(t)} -  \frac{{r^{\prime}}^2}{L(t^{\prime})}  \right) \right]
\nonumber \\
& & \times
\exp \left[ -\frac{i \hbar}{2\mu} \frac{x^2_{0n}}{a}\left( \frac{t}{L(t)} -  \frac{t^{\prime}}{L(t^{\prime})} \right) \right]
\nonumber \\
& & \times
j_0 \left( x_{0 n}  \frac{r}{L(t)} \right)~j_0 \left( x_{0 n}  \frac{r^{\prime}}{L(t^{\prime})} \right)~
\nonumber \\
& & \times
\frac{1}{\sqrt{4\pi}} \frac{1}{\sqrt{4\pi}}~,
\end{eqnarray}
where we have used $Y_{00} = 1/\sqrt{4\pi}$. The first two Bessel functions are
\begin{eqnarray}
j_0(x) &=& \frac{\sin x}{x}~, \nonumber\\
j_1(x) &=& \frac{\sin x}{x^2} - \frac{\cos x}{x} ~,
\end{eqnarray}
thus $x_{0n} = n \pi$ and $j_1(x_{0n}) = (-1)^{n+1}/n\pi$. Using these in eq. (\ref{eq: 1_1D_prop}), we find
\begin{eqnarray}
K(r, t ; r^{\prime}, t^{\prime}) &=& \frac{1}{4\pi} \frac{1}{rr^{\prime}} \frac{2}{\sqrt{L(t) L(t^{\prime})}}
\nonumber \\
& & \times
\exp \left[ \frac{i\mu u}{2\hbar} \left( \frac{r^2}{L(t)} -  \frac{{r^{\prime}}^2}{L(t^{\prime})}  \right) \right]
\nonumber \\
& & \times
\sum_{n=1}^{\infty}
\exp \left[ \frac{i \hbar}{2 \mu} \frac{n^2 \pi^2}{u} \left( \frac{1}{L(t)} - \frac{1}{L(t^{\prime})}  \right) \right]
\nonumber \\
& & \times
\sin \left( n\pi \frac{r}{L(t)} \right) \sin \left( n\pi \frac{r^{\prime}}{L(t^{\prime})} \right)~.
\end{eqnarray}
Now from eq. (\ref{eq: 1d-3d-prop}) we obtain
\begin{eqnarray}
K_{\text{1D}}(x, t ; x^{\prime}, t^{\prime}) &=& \frac{2}{\sqrt{L(t) L(t^{\prime})}}
\nonumber \\
& & \times
\exp \left[ \frac{i\mu u}{2\hbar} \left( \frac{x^2}{L(t)} -  \frac{{x^{\prime}}^2}{L(t^{\prime})}  \right) \right]
\nonumber \\
& & \times
\sum_{n=1}^{\infty}
\exp \left[ \frac{i \hbar}{2 \mu} \frac{n^2 \pi^2}{u} \left( \frac{1}{L(t)} - \frac{1}{L(t^{\prime})}  \right) \right]
\nonumber \\
& & \times
\sin \left( n\pi \frac{x}{L(t)} \right) \sin \left( n\pi \frac{x^{\prime}}{L(t^{\prime})} \right)~.
\end{eqnarray}
which is exactly eq. (32) of ref. \cite{LuCh-JPA-1992} for the
propagator of a particle in a 1D box. This equation
can be written in a compact form as a combination of $\vartheta_3$
functions \cite{Gr-PLA-1993}.
%
%


\section{Summary and Discussion}

In this letter we found solutions of the Schr\"{o}dinger equation
for a particle confined in a hard spherical trap with a moving
wall at constant velocity. We see in solutions
(\ref{eq: norm_waves}), except for the phase factor $\exp \left(-i
\int dt E_{l n m}(t) /\hbar \right)$ which has no coordinate
dependence, a coordinate-dependent phase $\exp \left[ \frac{i
\mu}{2\hbar} u \frac{r^2}{L(t)} \right]$ appears. It has been shown
that this factor leads to an effective quantum non-local interaction
with the boundary: even though the particle is nowhere near the
walls, it will be affected  \cite{Ma-JPA-1992, Mo-2011-quant}.

From fig. \ref{fig: B_Coefficients}, one sees that as the velocity
of the wall increases, larger amounts of energy states other than
the initial one, {\it{i.e.,}} $u_{11m}$, are mixed in. Fig.
\ref{fig: Energy_Expectation} shows that for rapid contraction,
energy expectation value increases faster than the $1/L^2(t)$
increase which would be obtained in a quasistatic contraction. These
results are in agreement with the ones of ref. \cite{DoRi-AJP-1969}
obtained for a particle in an infinite square well with one wall in
uniform motion. Confinement of the particle to a smaller region
leads to enhancement of the energy expectation value. This can be
explained by an application of the "old quantum theory"
\cite{Pi-AJP-1989} or by uncertainty relations \cite{Wi-JPA-1983}.

In the process of expansion, there are two
characteristic times involved: $t_{\text{e}}$, over which the
parameters of the system change appreciably, and $t_{\text{i}}$,
representing the motion of the system itself. In our calculations,
$t_{\text{e}} = a/u$ and $t_{\text{i}} = a/v_{ln}$. Figure \ref{fig:
wave_r} shows that for $t_{\text{e}} \gg t_{\text{i}}$ ($u \ll
v_{ln}$), the particle, initially in the state $u_{0,5,0}$, will end
up in the corresponding state of the expanded well. This process
characterizes an adiabatic one for which external conditions change
gradually \cite{Gr-book-1994}. While, in the opposite limit, rapidly
changing conditions prevent the system from adapting its
configuration during the process, hence the probability density
remains almost unchanged.

Noticing fig. \ref{fig: wave_t}, one sees a quasi-classical behavior
in the high-energy limit \cite{Go-PB-2007} as the velocity of the
wall increases. A non-monotonous increasing behavior of the density
is seen for $T < T_1$ only when $u > v_{ln}$, while for $T > T_2$ a
non-monotonous decreasing behavior is seen irrespective of the wall
velocity. These results are in contrast to classical mechanics. The
height of the first maximums decrease with $u$. The constructive
interference with the reflected components from the wall for $u <
v_{ln}$ leads to this enhancement. Long time behavior of the density
in the given observation point, is the same for all values of the
wall velocity, which is not an unexpected result noticing the
behavior of functions $\Psi_{lnm}$ at long times.

Propagator of the problem was derived using the spectral decomposition.
\\
\\

{\bf{Acknowledgment}}
The author would like to acknowledge two anonymous referees for valuable comments on an earlier draft of the paper.
Financial support from the University of Qom is gratefully acknowledged.


\end{document}